\documentclass[aps,prl,twocolumn,superscriptaddress,10pt]{revtex4-1}

\usepackage[latin1]{inputenc}
\usepackage{graphicx}
\usepackage{amsmath}
\usepackage{enumitem}

\usepackage{times}

\begin{document}

\title{Dynamics of the prey prehension by chameleons through viscous adhesion: A multidisciplinary approach}

\author{Fabian Brau}\email{fabian.brau@ulb.ac.be}\thanks{Present address: Nonlinear Physical Chemistry Unit, Facult\'e des Sciences, Universit\'e Libre de Bruxelles (ULB), CP231, 1050 Brussels, Belgium}
\affiliation{Laboratoire Interfaces \& Fluides complexes, Universit\'e de Mons, B-7000 Mons, Belgium}
\author{D\'eborah Lanterbecq}
\affiliation{Laboratoire Interfaces \& Fluides complexes, Universit\'e de Mons, B-7000 Mons, Belgium}
\author{Le\"ila-Nastasia Zghikh}
\affiliation{Laboratoire d'histologie, Universit\'e de Mons, B-7000 Mons, Belgium}
\author{Vincent Bels}
\affiliation{Institut des biosciences, Universit\'e de Mons, B-7000 Mons, Belgium and UMR 7205 CNRS/MNHN-UPMC-EPHE, Mus\'eum national d'Histoire naturelle, 55 rue Buffon, F-75005 Paris Cedex 5, France}
\author{Pascal Damman}\email{pascal.damman@umons.ac.be}
\affiliation{Laboratoire Interfaces \& Fluides complexes, Universit\'e de Mons, B-7000 Mons, Belgium}




\begin{abstract}
Chameleons are able to capture very large preys by projecting the tongue and retracting it once it is in contact with preys. A strong adhesion between the tongue tip and the prey is therefore required during the retraction phase to ensure a successful capture. To determine the mechanism responsible for this strong bond, the viscosity of the mucus produced at the chameleon's tongue pad is measured by using the viscous drag exerted on rolling beads by a thin layer of mucus. The viscosity of this secretion is found to be about 400 times larger than the one of human saliva. With a dynamical model for viscous adhesion describing the motion of the compliant tongue and of the prey during the retraction phase, the evolution of the maximum prey size with respect to the chameleon body length is derived. This evolution is successfully compared with in vivo observations for various chameleon species and shows that the size of the captured preys is not limited by viscous adhesion thanks to the high mucus viscosity and the large contact area between the prey and the tongue. 
\end{abstract}

\date{\today}

\maketitle

Chameleons are ambush opportunistic predators feeding on a large variety of invertebrate and vertebrate animals ranging from ants to birds and lizards~\cite{schwenk}. They remain motionless, hidden from their own predators, and wait for active preys to come within the reach of a strike. They have developed a highly specialized feeding system based on the ballistic projection of the tongue as far as $1-2$ body length with accelerations of up to 500 m/s$^{2}$~\cite{wainwright91,wain92a,wain92b,degroot04} combined with a very efficient adhesion allowing the capture of preys weighing up to 30\% of their own weight~\cite{herr09}. The ability to project the tongue with such high acceleration has been fairly well understood~\cite{degroot04} but the dynamics of the capture and the mechanism responsible for the strong adhesion between the prey and the tongue remains unclear. Interlocking, where the roughness of both prey and tongue surfaces self-adjust to make physical crosslinks~\cite{schwenk,higham13}, and suction mechanism, similar to the one at play in rubber suction pads~[4], have been proposed to supplement viscous (Stefan) adhesion~\cite{stefan,bikerman}. Here, we propose that viscous adhesion alone could explain the outstanding ability of chameleons to capture large preys without resorting to other mechanisms.

\begin{figure}
\centerline{\includegraphics[width=\columnwidth]{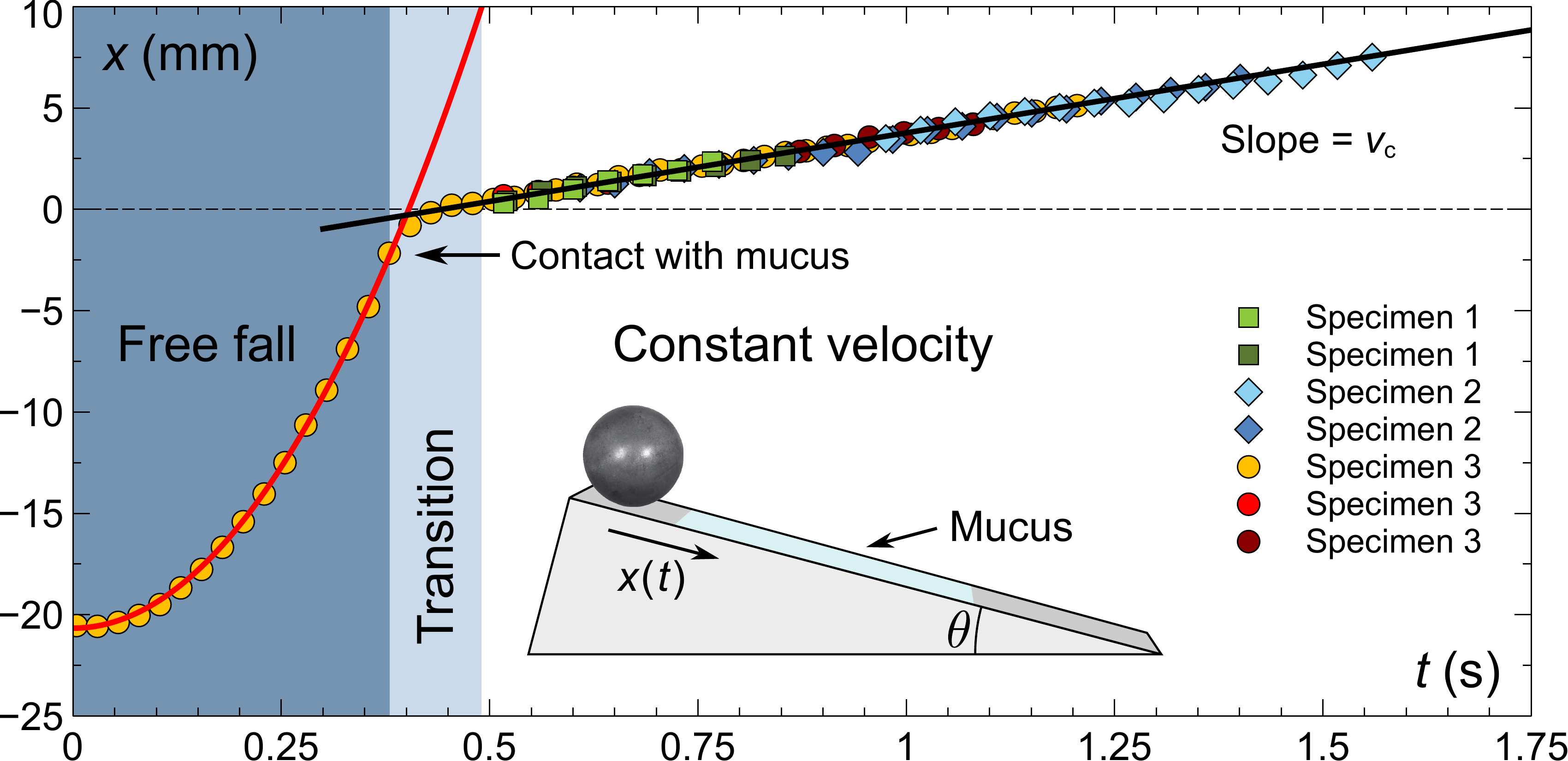}}
\caption{Position as a function of time of a spherical steel bead rolling down an inclined plane covered by a mucus layer of thickness $h_s$ for three specimens of {\it Chamaeleo calyptratus}. A free fall motion, shown only for one sequence for clarity, is followed by a regime at constant velocity, $v_c = (6.58 \pm 0.06)$ mm/s, once the bead is in contact with the fluid. The mass of the bead is $m_b=0.175$~g ($R=1.75$~mm and $\rho_b=7795$~kg/m$^3$).}
\label{beadmotion}
\end{figure}

The viscosity of the mucus secreted at the chameleon's tongue pad is a crucial parameter to study the adhesion mechanism. However, this specific fluid is produced in very small amounts by glands in the tongue pad~\cite{schwenk} and its viscosity remains today completely unknown. To overcome this difficulty, the drag exerted by viscous forces on small steel beads rolling on a thin film of mucus is used to measure indirectly the viscosity~\cite{bico09}. The mucus is collected from the contact between the tongue pad and a microscope slide placed in front of a prey to provoke a shoot of the tongue. The slide is then placed without delays on a support tilted with an angle~$\theta$. The motion of a bead rolling down the slide is then recorded immediately with a camera (250 fps). As shown in Fig.~\ref{beadmotion}, after a transient stage, the bead moves at constant velocity, $v_c$, which is determined by $\theta$, the fluid and the bead properties as follows
\begin{equation}
\label{V-bico}
v_c = C_v \, (\gamma/\eta)\, (\sin \theta)^{\alpha} \left(\rho_{b} g R^2/\gamma\right)^{\beta} \left(R/h_s\right)^{1/2}.
\end{equation}
The quantities $\gamma = 70$ mN/m, $\eta$ and $h_s$ are the surface tension, viscosity and thickness of the mucus layer~\cite{rem-gamma}. $\rho_b$ and $R$ are the density and radius of the bead. $g$ is the gravitational acceleration, $\alpha=1.6 \pm 0.06$ and $\beta =1.35 \pm 0.05$ are numerical constants determined experimentally~\cite{bico09}. $C_v=0.014$ is the relevant value to be used in our case because the capillary number ($\eta v_c/\gamma \sim 0.04$) is much smaller than~1~\cite{bico09}. The validity of the method was checked with fluids of known viscosity~\cite{sup_mat}.
The various measurements of the bead motion reported in Fig.~\ref{beadmotion} yield comparable velocities indicating that the fluid parameters, and in particular $h_s$, were similar in all experiments. From Eq.~(\ref{V-bico}), the viscosity is found to lie in the range $\eta = 0.4 \pm 0.1$~Pa~s which is much larger than the one of human saliva ($\sim 10^{-3}$~Pa~s~\cite{briedis}). The relatively large error is related to the uncertainty about the thickness of the mucus layer, $h_s=25 \pm 10$~$\mu$m, that was determined by weighing and measuring the area of the film. Notice that the film thickness $h_0$ involved during a capture (see below) is larger than $h_s$ since only a part of the mucus is deposited on the microscope slide once the tongue is detached.

\begin{figure}[t]
\centerline{\includegraphics[width=\columnwidth]{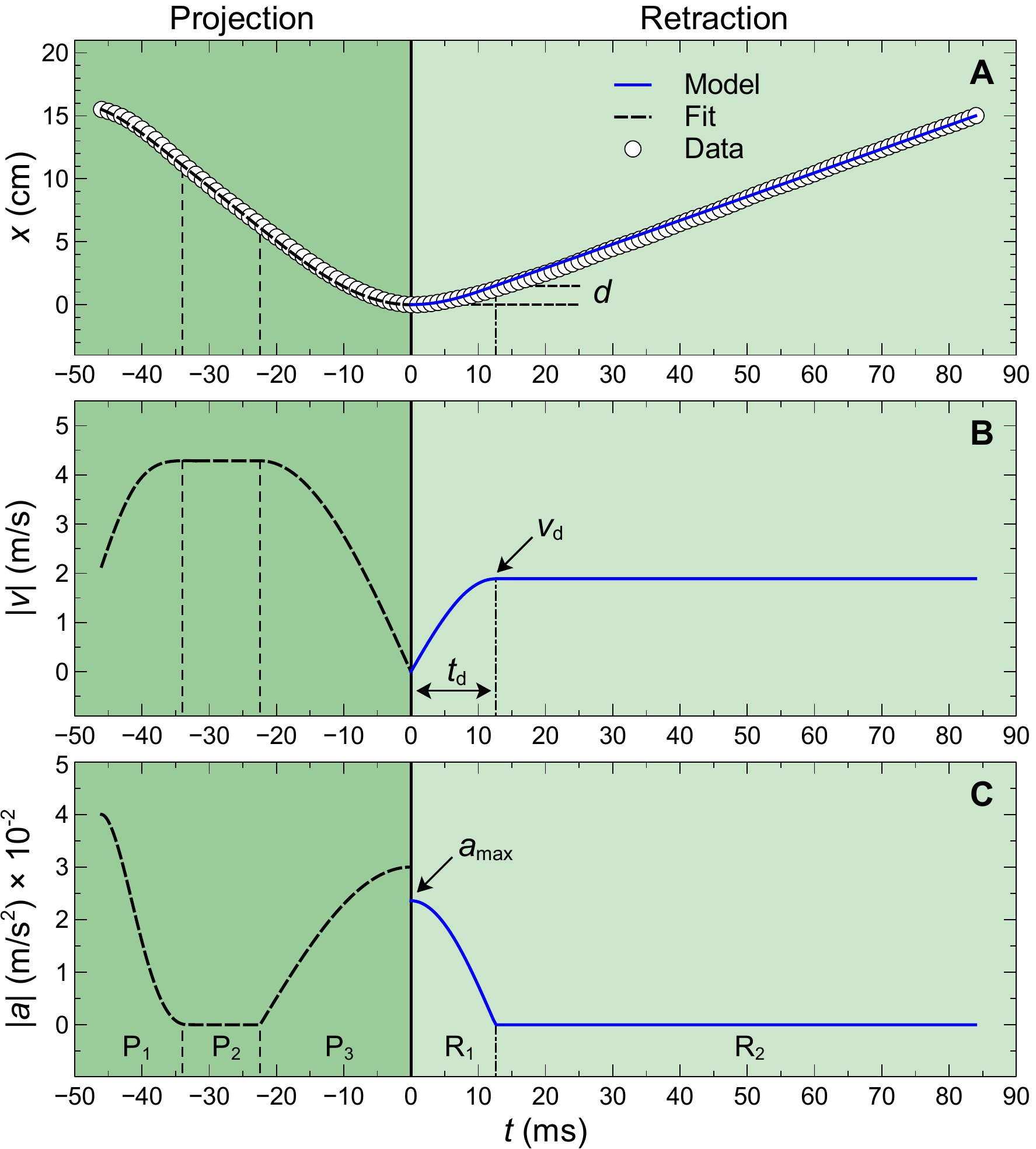}}
\caption{Kinematics profiles for one representative capture of a cricket ($m_p\simeq0.5$ g) by a {\it Chamaeleo calyptratus} specimen ($L_{\text{SVL}} = 15$ cm). ({\bf A}) Position of the tip of the chameleon tongue as a function of time measured with respect to the onset of retraction. The evolution predicted by the model (\ref{model}) is also shown where the mean value of all parameters are used together with $k = 212 \, L_{\text{SVL}}$ and $d=0.1 \, L_{\text{SVL}}$). The projection phase has been fitted qualitatively by an ad hoc function to illustrate the relevant regimes discussed in the text. $d$ is the distance over which the acceleration is significant during the retraction phase. ({\bf B})-({\bf C}). Velocity and acceleration as a function of time obtained from the model (retraction) and the fit (projection) where some relevant quantities of the retraction phase are indicated. The vertical dashed lines define various regimes of the prey capture (see text).}
\label{figcine}
\end{figure}

This unexpectedly large mucus viscosity strongly suggests that the prey very likely sticks to the chameleon's tongue through viscous adhesion. The value of this quantity is however not sufficient to determine the adhesion strength. 
Viscous forces are determined by the viscosity together with the flows within the fluid layer~\cite{stefan,bikerman}. The strain/shear rate in the fluid film, which is only significant during the retraction phase, should also be known to determine the magnitude of the viscous forces. 
For this purpose, we introduce a dynamical model for the retraction phase.

Figure~\ref{figcine} shows typical kinematics data of a prey capture recorded with a high speed camera (1000 fps) featuring two main phases: tongue projection and retraction. The chameleon first estimates the distance while the tongue slowly protrudes out of the jaws~\cite{harkness}. Then, the accelerator muscle contracts radially to squeeze against the entoglossal process of the hyoid skeleton which leads to the projection of the tongue with a high acceleration~\cite{degroot04,mull04,anderson10}. The acceleration sharply decreases (P$_{1}$ regime) to almost vanish such that the tongue moves at a roughly constant velocity (P$_{2}$ regime)~\cite{wain92a,degroot04}. The tongue then decelerates, its elongation approaching the maximal extension, and hits the prey before to stop when the velocity vanishes and the acceleration reaches a local extremum (P$_{3}$ regime). The tongue retraction then starts, with a retraction force of roughly $1$ N~\cite{herrel01}, the velocity increases while acceleration decreases and finally almost vanish (R$_{1}$ regime). The rest of the tongue retraction is performed at essentially constant velocity (R$_{2}$ regime). 

This typical sequence shows that, after the tongue whipped out the mouth at high acceleration, it moves essentially at roughly constant velocity except near the retraction point~\cite{degroot04,herrel00}. Therefore, the tongue does not behaves like a stretched elastic material over its whole elongation but only in a small region around the capture point. This is consistent with the observation that the tongue is made of nested sheaths sliding along one another like the tubes of an extending telescope~\cite{degroot04}. Stretching only occurs once this telescopic structure has been fully deployed. To describe the retraction phase, we thus model the tongue as a spring of stiffness $k$ stretched over a distance $d$ which is only a fraction of the prey-jaws distance (see Ref.~\cite{sup_mat} for the case of a constant retraction force). 

\begin{figure*}[t]
\centerline{\includegraphics[width=\textwidth]{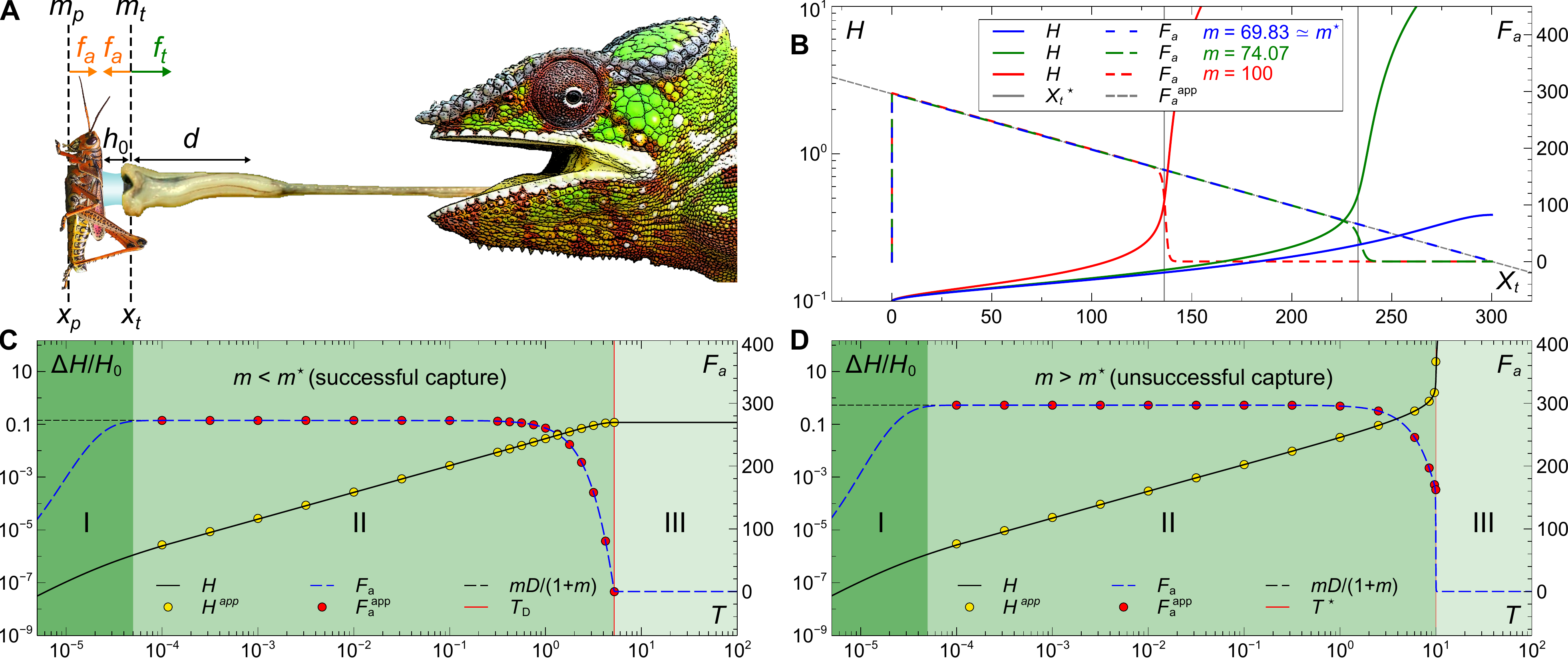}}
\caption{(color online). ({\bf A}) Schematic of the initial state of the retraction phase with the adhesion force $f_a$, the applied force $f_t$, the initial thickness of the mucus layer $h_0$ and the distance $d$ over which the retraction force applies. ({\bf B}) Evolution of the normalized mucus thickness $H$ and adhesive force $F_a$ as a function of the normalized position of the tongue tip $X_t$ obtained by solving Eqs.~(\ref{eq-motion}) numerically for $H_0=0.1$, $D = 300$ and several values of $m=m_p/m_t$. ({\bf C})-({\bf D}) Evolution of the relative variation of $H$ and $F_a$ as function of the normalized time $T$ for $H_0=0.1$, $D = 300$ and $m=10$ (C) or $m=100$ (D) showing three regimes: (I) onset of retraction, (II) retraction without detachment ($\Lambda=2.76$) (C) or with detachment ($\Lambda=0.84$) (D) and (III) inertial motion. The approximate expressions (\ref{adh-p-2}), (\ref{Fa-app}) and (\ref{H-app}) for $F_a$ and $H$ are also shown.}
\label{figforce}
\end{figure*}

The retraction phase is described by considering a prey of mass $m_p$ at a position $x_p$ attached by a viscous fluid of thickness $h$ to a tongue of mass $m_t$ and position $x_t$. A force $f_t$ retracts the tongue and produces an increase of the mucus thickness ($dh/dt \equiv \dot{h} >0$) inducing a viscous force $f_a$ acting on both the tongue and the prey, see Fig.~\ref{figforce}A. This force is due to the Poiseuille flow that is created in an incompressible viscous fluid when it deforms: $f_a = ({3}/{2 \pi}) {\eta \Omega^2 \dot{h}}h^{-5}$, where $\Omega$ is the (constant) fluid volume~\cite{stefan,bikerman,rem1}. The equations of motion are~\cite{rem2}
\begin{subequations}
\label{model}
\begin{align}
m_t \ddot{x}_t &= f_t - f_a \nonumber \\
&= k(d+h_0-x_t)\, {\cal H}(d+h_0-x_t)- \alpha \dot{h}h^{-5}, \\
\label{prey-eq}
m_p \ddot{x}_p & = f_a =\alpha \dot{h}h^{-5}, \quad \alpha =3\eta \Sigma^2 h_0^2/2\pi, \\
h &= x_t-x_p,
\end{align}
\end{subequations}
with the initial conditions $x_p(0)= \dot{x}_p(0)= \dot{x}_t(0)=0$ and $x_t(0)=h_0$. $\Sigma=\Omega/h_0$ is the initial contact area between the prey and the tongue and $h_0$ is the initial mucus thickness. The Heaviside function ${\cal H}(x)$ is added to describe an applied (retraction) force vanishing when the tongue has reached a retraction distance $d$ (\textit{i.e.} when $x_t>d+h_0$). Therefore, beyond that distance, only inertia is involved, $h$ stays constant and both the tongue and the prey move at constant speed. This implies that the prey can only detach before the tongue reaches the position $d + h_0$.  

This system is characterized by the length scale $\ell=(\alpha^2/m_t k)^{1/10}$ and the time scale $\tau=(m_t/k)^{1/2}$ which are used to adimensionalize the equations of motion as follows
\begin{subequations}
\label{eq-motion}
\begin{align}
\label{tongue}
\ddot{X}_t &= (D+H_0-X_t)\, {\cal H}(D+H_0-X_t)- \dot{H} H^{-5}, \\
\label{prey}
\ddot{X}_p &= m^{-1} \, \dot{H} H^{-5}, \quad m = m_p/m_t, \\
\label{H}
H &= X_t-X_p.
\end{align}
\end{subequations}
To illustrate the dynamics produced by this model, Fig.~\ref{figforce}B shows the evolutions of $H$ and $F_a=\dot{H}/H^{5}$ as a function of the tongue position $X_t$ for some typical values of $H_0$ and $D$. Depending on the magnitude of $m$, two different behaviors are observed. When $m\le m^{\star}$, $H$ stays finite and there is no detachment. The adhesive force $F_a$ always vanishes precisely at $X_t =D+H_0$ where $F_t$ cancels and $H$ is therefore constant beyond that distance ($\dot{H}=0$). The prey stays then attached to the tongue since both move at the same speed. However, when $m> m^{\star}$, $H$ diverges at some detachment distance $X_t^{\star}$ leading to a sudden fall of the adhesive force. Therefore, it exists a critical value $m=m^{\star}(H_0,D)$ above which detachment occurs. We derived this relation analytically below.

Except from a tiny region near the onset of retraction, Fig.~\ref{figforce}B shows that when the adhesive force is non vanishing it satisfies in good approximation the following relation
\begin{equation}
\label{adh-p-2}
F_a \simeq F_a^{\text{app}} = m(D+H_0-X_t)/(1+m).
\end{equation}
Substituting Eq.~(\ref{adh-p-2}) into Eqs.~(\ref{tongue}) and (\ref{prey}) and solving the ODE leads to the tongue and prey positions
\begin{subequations}
\begin{align}
\label{xt}
X_t &\simeq H_0+D\, (1-\cos \omega T), \\
\label{xp}
X_p &\simeq D\, (1-\cos \omega T), \quad \omega = (1+m)^{-1/2}.
\end{align}
\end{subequations}
These approximates solutions imply that $H$ stays close to $H_0$ during the retraction. The correct evolution of $H$ can nevertheless be obtained from the adhesive force which, using Eqs.~(\ref{adh-p-2}) and (\ref{xt}), reads
\begin{equation}
\label{Fa-app}
F_a \simeq F_a^{\text{app}} = \frac{\dot{H}}{H^5}\simeq \frac{D \, m}{1+m} \cos \omega T.
\end{equation}
An integration leads to the evolution of the mucus thickness as a function of time
\begin{equation}
\label{H-app}
H \simeq H^{\text{app}} = H_0\left[\frac{\Lambda}{\Lambda-\sin\omega T}\right]^{1/4}, \quad \Lambda = \frac{\sqrt{1+m}}{4 mD H_0^4}.
\end{equation}
These approximate expressions (\ref{Fa-app})-(\ref{H-app}) agree very well with the numerical solutions as seen in Fig.~\ref{figforce}C-D and are valid until $F_a$ vanishes.
Interestingly, the approximate solution captures very well the divergence of $H$ that occurs only when $\Lambda \le 1$. In such a case (unsuccessful capture), the adhesive force vanishes while the retractive force still applies to the tongue leading to a prey detachment at time $T=T^{\star}$, or equivalently at $X_t=X_t^{\star}$, given by
\begin{equation}
T^{\star}=\frac{1}{\omega}\arcsin\Lambda, \quad X_t^{\star} = H_0+ D\left(1-\sqrt{1-\Lambda^2}\right).
\end{equation}
After detachment, the prey moves at constant speed $V_p(T^{\star})=(4m H_0^4)^{-1}$ while the tongue still accelerates until it reaches the position $D+H_0$, where $F_t$ vanishes.

When $\Lambda >1$, the mucus thickness $H$ increases to saturate at a finite value $H = H_0 [\Lambda/(\Lambda-1)]^{1/4}$, see Eqs.~(\ref{xt}), (\ref{H-app}) and Fig.~\ref{figforce}C. Therefore, the adhesive force acts on the prey until the retractive force vanishes when the tongue reaches the position $D+H_0$ at time $T=T_D=\pi/(2\omega)$. 

The necessary condition for a successful capture can be written as $\Lambda >1$. Considering the definition of $\Lambda$, it imposes the following constraint on the prey mass: $m_p < m_p^{\star}=m_t/(16 D^2 H_0^8)$, assuming $m^{\star} \gg 1$. Returning to the physical variables, the maximum prey mass is therefore given by
\begin{equation}
\label{masse}
m_p^{\star} = \rho V^{\star} = \frac{9}{64\pi^2} \frac{\eta^2 \Sigma^4}{k d^2 h_0^4},
\end{equation}
where $\rho\simeq 1050$ kg/m$^3$ is a typical prey density and $V^{\star}$ the prey volume~\cite{matt08}. The values of the parameters assumed to be constant are: $\eta=(0.4\pm0.1)$~Pa~s and $h_0 \simeq 2 h_s = (50\pm 10)$~$\mu$m. In contrast, the morphological parameters, $k$, $\Sigma$ and $m_t$ depend on the snout-vent length, $L_{\text{SVL}}$, of the specimen. From kinematics and morphological data found in literature for various chameleons~\cite{sup_mat}, we get $k = (223 \pm 60) \, L_{\text{SVL}}$, $\Sigma = (4.8 \pm 1.2)\, 10^{-3} L_{\text{SVL}}^2$ and $m_t=(0.45\pm 0.09)\, L_{\text{SVL}}^3$ in MKS units. Therefore, we obtain the following order of magnitudes for the characteristic length and time scales: $\ell \simeq$ $0.3-0.5$ mm and $\tau\simeq$ $2-9$ ms for $L_{\text{SVL}}= 50-200$ mm. We also expect $d$ to scale linearly with $L_{\text{SVL}}$, $d=(0.2\pm 0.1)\, L_{\text{SVL}}$~\cite{sup_mat,rem3}. 
%

\begin{figure}
\centerline{\includegraphics[width=\columnwidth]{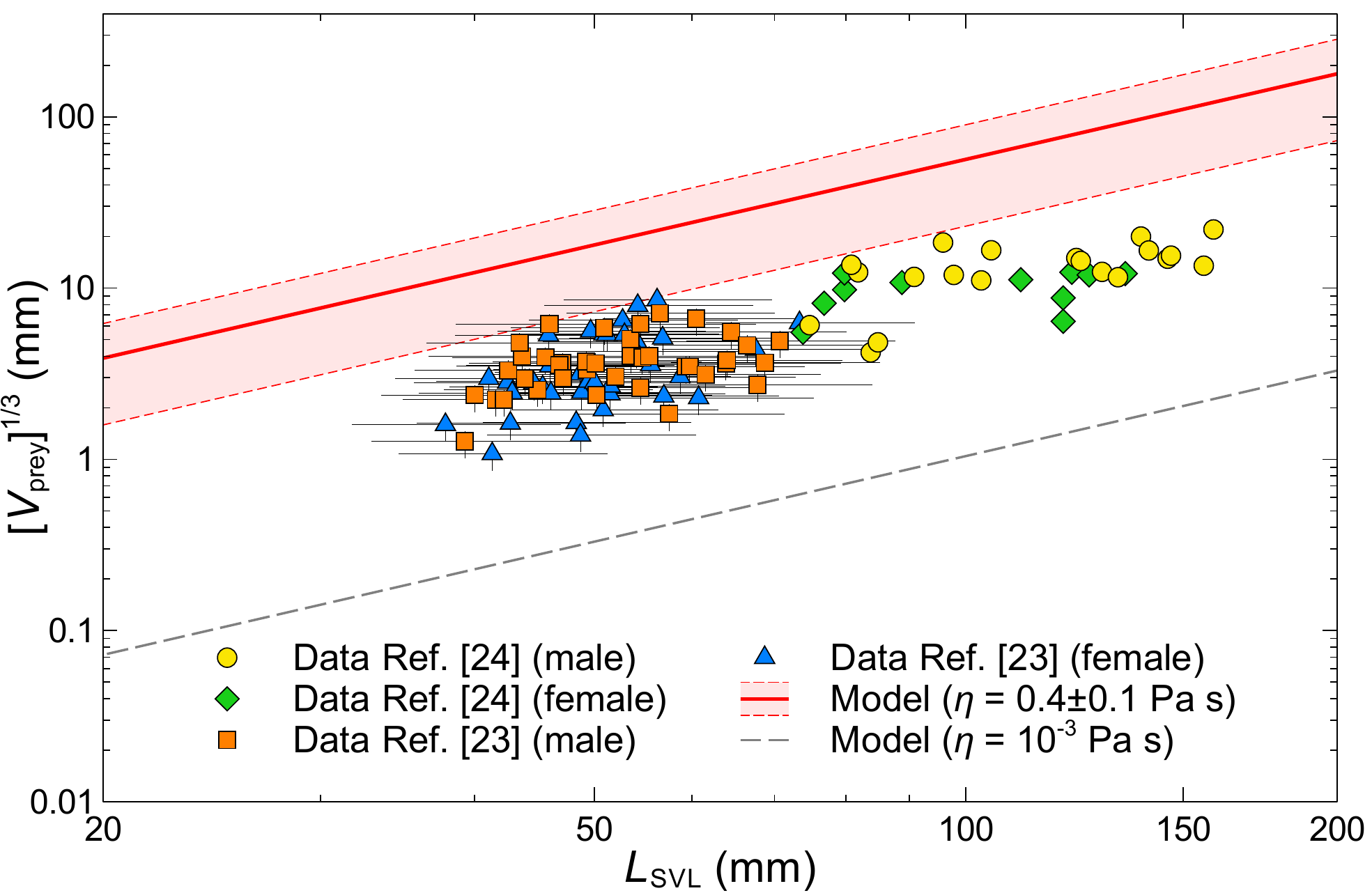}}
\caption{Measured maximum prey size as a function of $L_{\text{SVL}}$ obtained for a large sampling of chameleons~\cite{measey11,kraus12} plotted together with the prediction of a pure viscous adhesion model described by Eq.~(\ref{prey-size}). The theoretical evolution with the viscosity of human saliva is shown for comparison.}
\label{figmass}
\end{figure}

Using these parameter values, Eq.~(\ref{masse}) can be written as a function of the chameleon body size as
\begin{equation}
\label{prey-size}
{V^{\star}}^{1/3} = (2.7 \pm 1.6)\, L_{\text{SVL}}^{5/3},
\end{equation}
in MKS units. 
In the literature, two studies report in vivo analysis of the stomachal contents of a large number of chameleons to determine the mean maximum prey size in relation to the chameleon SVL~\cite{measey11,kraus12,sup_mat}. Figure~\ref{figmass} shows a comparison between these data and Eq.~(\ref{prey-size}). The maximum prey size estimated from the adhesion model is close but always larger than the experimental data and follows the global trend of the data. Viscous adhesion alone is therefore largely sufficient to allow the capture of very large preys. Actually, the adhesive mechanism appears to be outsized with respect to the usual preys found in stomachal contents (see Fig.~\ref{figmass}). This outstanding adhesion strength allows however chameleons to capture birds, lizards or mammals when they have the opportunity. Viscous forces could also explain the reported capture of prey weighing $30 \%$ of the chameleon body mass~\cite{herr09}. For a specimen with $L_{\text{SVL}}=0.1$ m, for example, the lower estimate of the maximum prey mass is about $60 \%$ of the chameleon body mass. Notice that during a capture, the maximum prey size could be lowered by a smaller contact area due to an imperfect shooting or by the gripping of the prey which are not taken into account here. Some ``safety factor" is thus necessary to ensure a successful capture even when the conditions are not optimal. 

Considering the maximum retraction force $f_{max} = kd$, Eq.~(\ref{masse}) can be rewritten as $m_p^{\star} \sim k \eta^2 \Sigma^4 /f_{max}^2 h_0^4$. It may appear paradoxical that $m_p^{\star}$ increases when $f_{max}$ decreases. Actually, the prey is attached to the tongue by a fragile viscous bound. If a large retraction force is applied to an heavy prey, the inertial force will be so large that the bound will break. Instead, if a small retraction force is applied, the viscous bound may resist to gentle displacement of the prey, \textit{i.e.} with a very small acceleration. 
In fact, the fast kinematics essential to capture a prey at long distance without allowing it to dodge the tongue imposes that the retraction force cannot be arbitrarily small which limits the maximum prey mass.

Equation (\ref{masse}) shows that two parameters influence positively the adhesive trap: the fluid viscosity and the contact area. The specific shape of the tip of the chameleon tongue which is large and form a kind of cup during the capture due to the action of specific muscles~[4] allows a drastic increase of the tongue-prey contact area, $\Sigma$. Indeed, a cup shape with highly deformable lips allows a large embedding of the prey within the tongue. Interestingly, \textit{Plethodontidae} (salamanders) use also a ballistic tongue to capture preys and have also developed a large and flexible tongue pad to engulf the prey and thus maximize the contact area~\cite{deban97,deban06}. Experiments to measure the viscous adhesion strength against other possible adhesion mechanisms should therefore be conducted very carefully with a constant contact area. A large mucus viscosity, as measured above, is also essential. For instance, if the viscosity of the chameleon tongue secretion was similar to human saliva ($\eta\simeq 10^{-3}$ Pa~s), Eq.~(\ref{masse}) shows that maximum prey mass would be reduced by roughly a factor of $10^5$ making this mode of capture extremely inefficient. 

\vspace{0.2cm}
\begin{acknowledgments}
\noindent {\bf Acknowledgments:} The lizard specimens were provided by C. Remy (Mus\'ee d'Histoire Naturelle de Tournai). The authors acknowledge C. Gay and D. Nonclercq for fruitful discussions. This work was partially supported by the MECAFOOD ARC research project from UMONS. F.B. acknowledges financial support from the Government of the Region of Wallonia (REMANOS Research Programme).
\end{acknowledgments}

\clearpage

\begin{center}
{\bf Dynamics of the prey prehension by chameleons through viscous adhesion: A multidisciplinary approach\\{\small \textit{Supplementary materials}}}
\end{center}

\section{Materials and methods}

\noindent \textbf{\textit{Biological organisms}} -- For the experiments, we used 5 chamelons (\textit{Chamaeleo calyptratus}). \vspace{2mm} \\
\textbf{\textit{Rolling beads experiments}} -- The viscosity of the adhesive fluid was measured from a rolling bead method. Steel beads were used with known mass ($0.175$ g) and diameter ($3.5 $ mm). The mucus is collected from the contact between the tongue pad and a microscope slide placed in front of a prey to provoke a shoot of the tongue. The thickness of the fluid is measured by weighing on a micro-balance and measuring the visible area of the film on a photography of the slide. The slide is then placed without delays on a support tilted with an angle of $10^{\circ}$ with respect to the horizontal. The motion of a bead rolling down the slide is then recorded immediately with a camera (250 fps). The position of the bead was obtained by direct analysis of digitalized video with ImageJ. The velocity of the bead was then measured in the regime where it is constant. The validity of the method was checked with fluids of known viscosity (silicone oil), see next section. \vspace{2mm} \\
\textbf{\textit{Kinematics of capture}} -- Sequences of cricket capture by Chamaeleo calyptratus were recorded with a Photron high speed camera operating at 1000 fps. The tongue pad position, $x(t)$, were directly measured with ImageJ on the movie.

\section{Rolling beads}

\begin{figure}
\centerline{\includegraphics[width=\columnwidth]{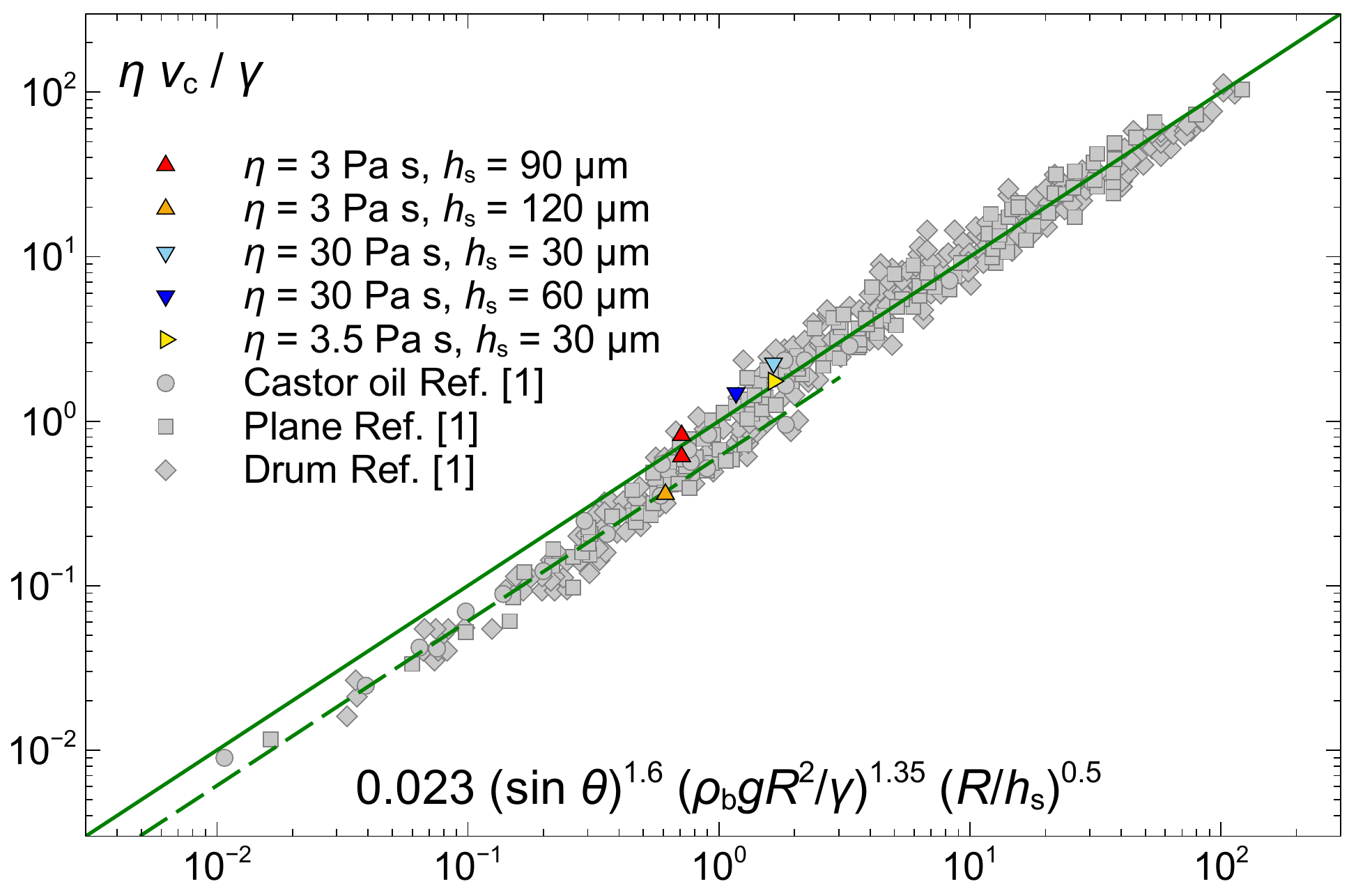}}
\caption{Evolution of the rescaled constant velocity $v_{\text{c}}$ as a function of the relevant combinaison of the control parameters. Data from Ref.~[1] are characterized by the following ranges for the physical parameters: $1.6$ mm $\le R \le 16$ mm, $1.4$ g/cm$^{3} \le \rho_{\text{b}} \le 14.9$ g/cm$^{3}$, $1$ mPa s $\le \eta \le 100$ mPa s and $0.02 \le \sin \theta \le 1$. For $\eta v_{\text{c}}/\gamma >1$, the scaling is given by Eq.~(1) of the main text with $D=0.023$. For $\eta v_{\text{c}}/\gamma <1$, the scaling is the same but with $D=0.014$. This scaling has been tested with our experimental setup by using three silicone oils of known viscosity ($\gamma = 20$ mN/m) for several thicknesses of the fluid layer as indicated and with $\theta = 10^{\circ}$.\label{fig-beads}}
\end{figure}

The drag exerted by viscous forces on small steel beads with known mass ($0.88$ g) and diameter ($6.0 $ mm) rolling on a fluid thin film is used to measure indirectly the viscosity of the mucus produced at the chameleon tongue tip~[1]. Our setup has been tested with three fluids (silicone oil) of known viscosity $\eta$. Glass plates are coated with those viscous oil using a bar coater to obtain a uniform fluid film with a given thickness $h_s$. A bead with a given radius $R$ and density $\rho_{\text{b}}$ is placed on the top of the coated glass plate titled with an angle $\theta$ and rolls down the slope. The position of the beads as a function of time is recorded with a high speed camera. After a transient stage, the bead moves at constant velocity, $v_{\text{c}}$, whose expression as a function of the various physical parameters of the system is given by Eq.~(1) of the main text. This expression has been intensively tested experimentally. Figure~\ref{fig-beads} shows the results of hundreds of experiments performed in Ref.~[1]. Therefore, we have only performed few experiments to verify that our setup was able to reproduce the expected results. As shown in Fig.~\ref{fig-beads}, our data are in good agreement with those reported in Ref.~[1]. 

\section{Relationships between relevant parameters and SVL}

\begin{figure}[t]
\centerline{\includegraphics[width=\columnwidth]{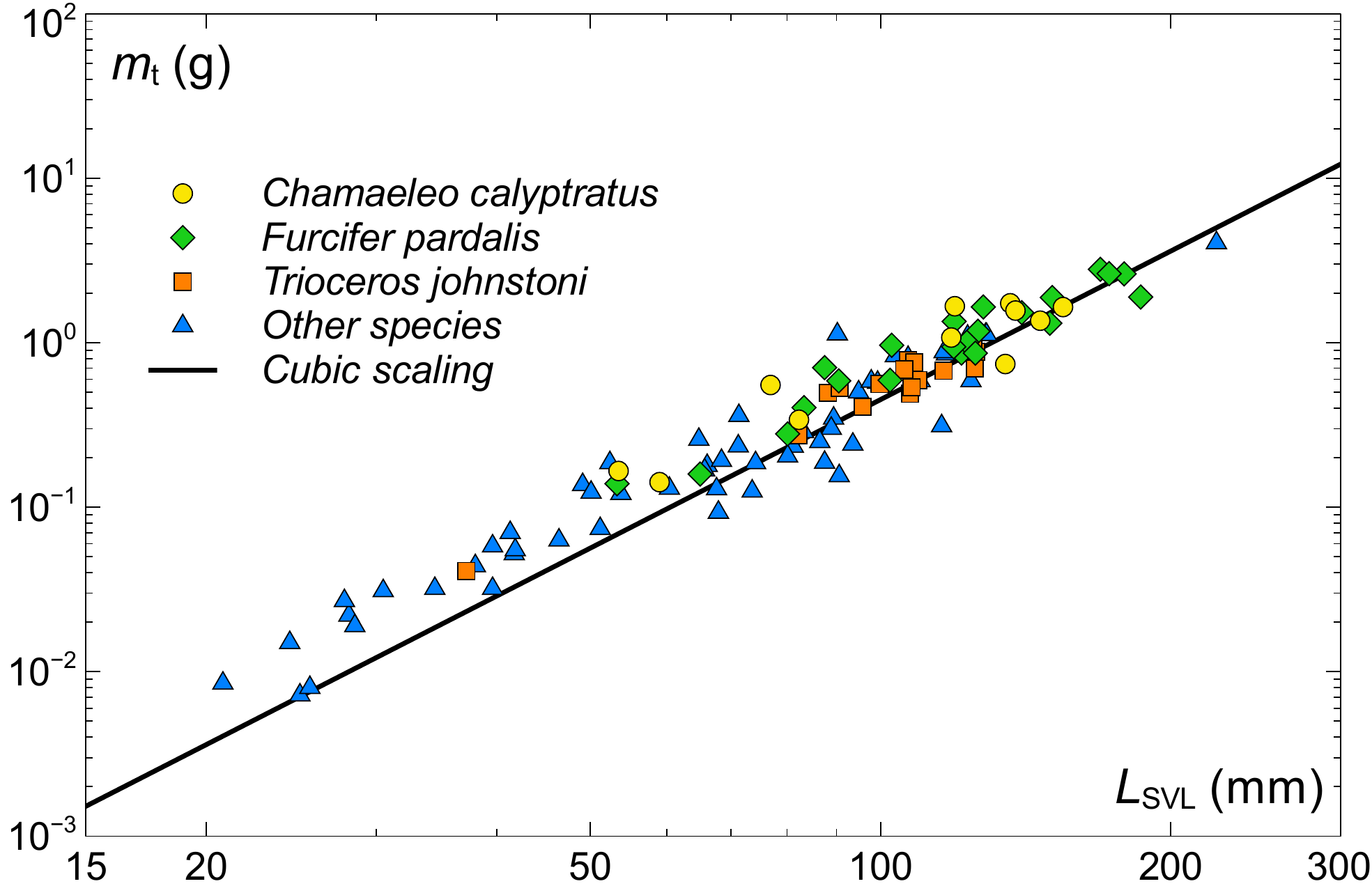}}
\caption{Evolution of the tongue mass, $m_t$, as a function of the SVL for 105 chameleons among 23 species from Ref.~[2].}
\label{fig-M_tongue}
\end{figure}

\noindent\textbf{\textit{Mass of the tongue}} -- One morphological parameter needed in the model developed in the main text is the mass of the tongue, $m_t$. The evolution of the tongue mass as a function of $L_{\text{SVL}}$ is shown in Fig.~\ref{fig-M_tongue} for a large number of specimens among various species~[2]. In good approximation, $m_t$ behaves as the third power of $L_{\text{SVL}}$ as follows
\begin{equation}
\label{m-tongue}
m_t = (0.45 \pm 0.09) \, L_\text{SVL}^3,
\end{equation}
in MKS units.

\begin{figure}[t]
\centerline{\includegraphics[width=\columnwidth]{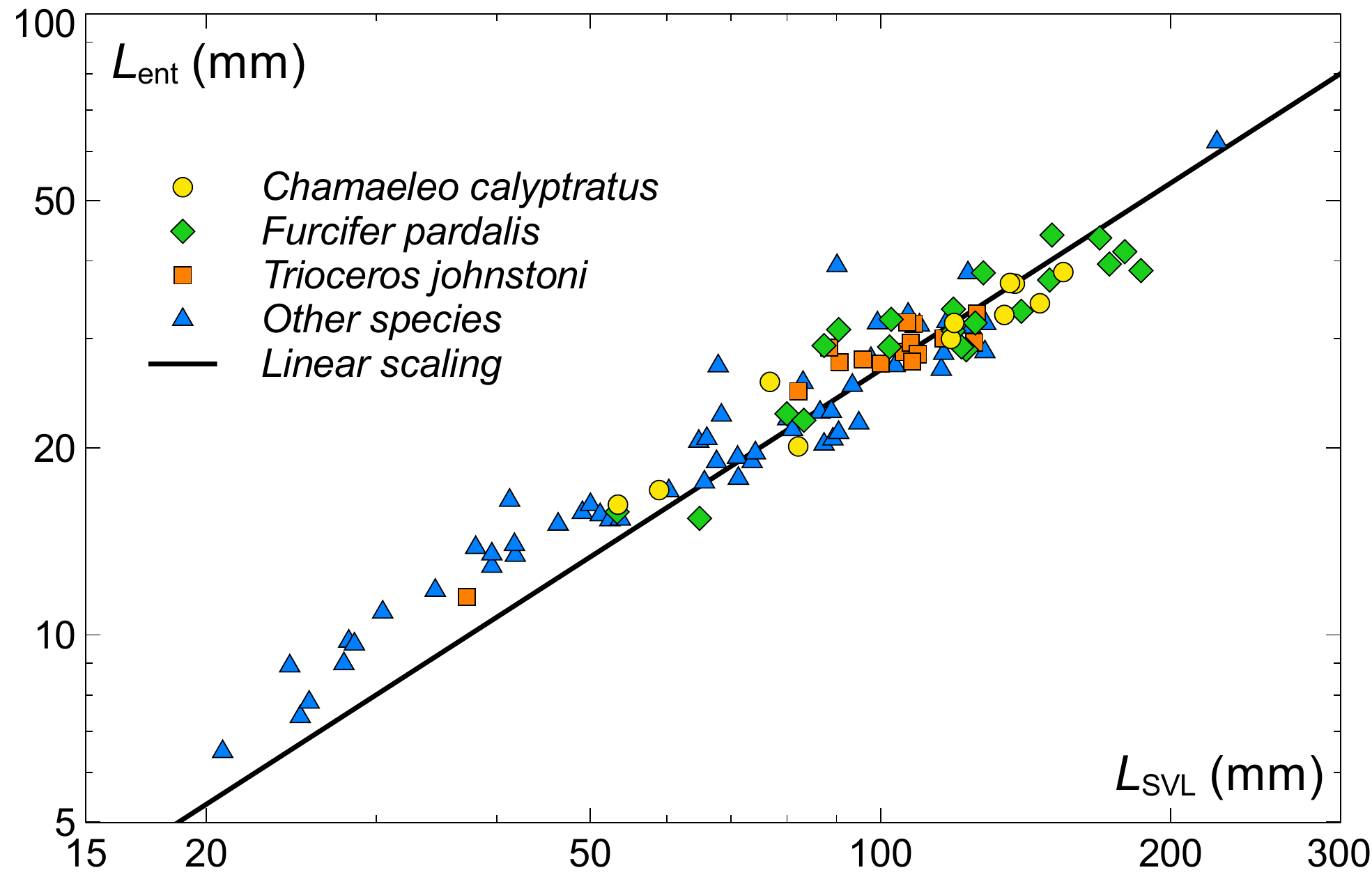}}
\caption{Evolution of the length of the entoglossal process, $L_{\text{ent}}$, as a function of the SVL for 105 chameleons among 23 species. The data come from Ref.~[2]. The unstretched tongue lengths, $L_t$, are approximated by $L_{\text{ent}}$.}
\label{fig-L_tongue}
\end{figure}

\noindent\textbf{\textit{Length and section of the tongue}} -- The length of the unstretched tongue, $L_t$, is approximated by the length of the entoglossal process (the tongue's skeletal support), $L_{\text{ent}}$, and is reported as a function of $L_{\text{SVL}}$ in Fig.~\ref{fig-L_tongue} for a large number of specimens among various species~[2]. In good accuracy, $L_t$ behaves linearly with $L_{\text{SVL}}$ as follows
\begin{equation}
\label{l-tongue}
L_t \simeq L_{\text{ent}} = (0.27 \pm 0.05) \, L_{\text{SVL}}.
\end{equation}
The cross section area of the tongue is obtained indirectly from the mass and the length of the tongue by approximating the volume of the unstretched tongue by the one of a cylinder, $m_t \simeq \rho_{\text{m}} S_t L_t$. Equations (\ref{m-tongue}) and (\ref{l-tongue}) together with the typical density of muscles ($\rho_{\text{m}} \simeq 1060$ kg/m$^3$)~[3], gives the evolution of the tongue section $S_t$ as a function of the chameleon size:
\begin{equation}
\label{s-tongue}
S_t = (1.6 \pm 0.4) \, 10^{-3} L_{\text{SVL}}^2.
\end{equation}

\noindent\textbf{\textit{Area of the tongue pad}} -- Another morphological parameter needed in the model developed in the main text is the contact area $\Sigma$ between the prey and the tongue. Since our intend is to quantify the maximum size of the prey captured by viscous adhesion, we consider that $\Sigma$ is given by the tongue size and not the prey size (chameleons are able to capture very large prey, such as birds or other lizards, having a size much larger than the tongue pad). If the tip of the tongue is approximated by half a sphere, then its area is twice the area of the tongue section $S_t$. Since the tip of the tongue deforms during a capture~[4], we consider that the contact area is roughly given by 3 times $S_t$. Therefore, using Eq.~(\ref{s-tongue}), we get the following scaling:
\begin{equation}
\label{sigma}
\Sigma \simeq 3 S_t \simeq (4.8 \pm 1.2) \, 10^{-3} \, L_{\text{SVL}}^2.
\end{equation}

\begin{figure}[t]
\centerline{\includegraphics[width=\columnwidth]{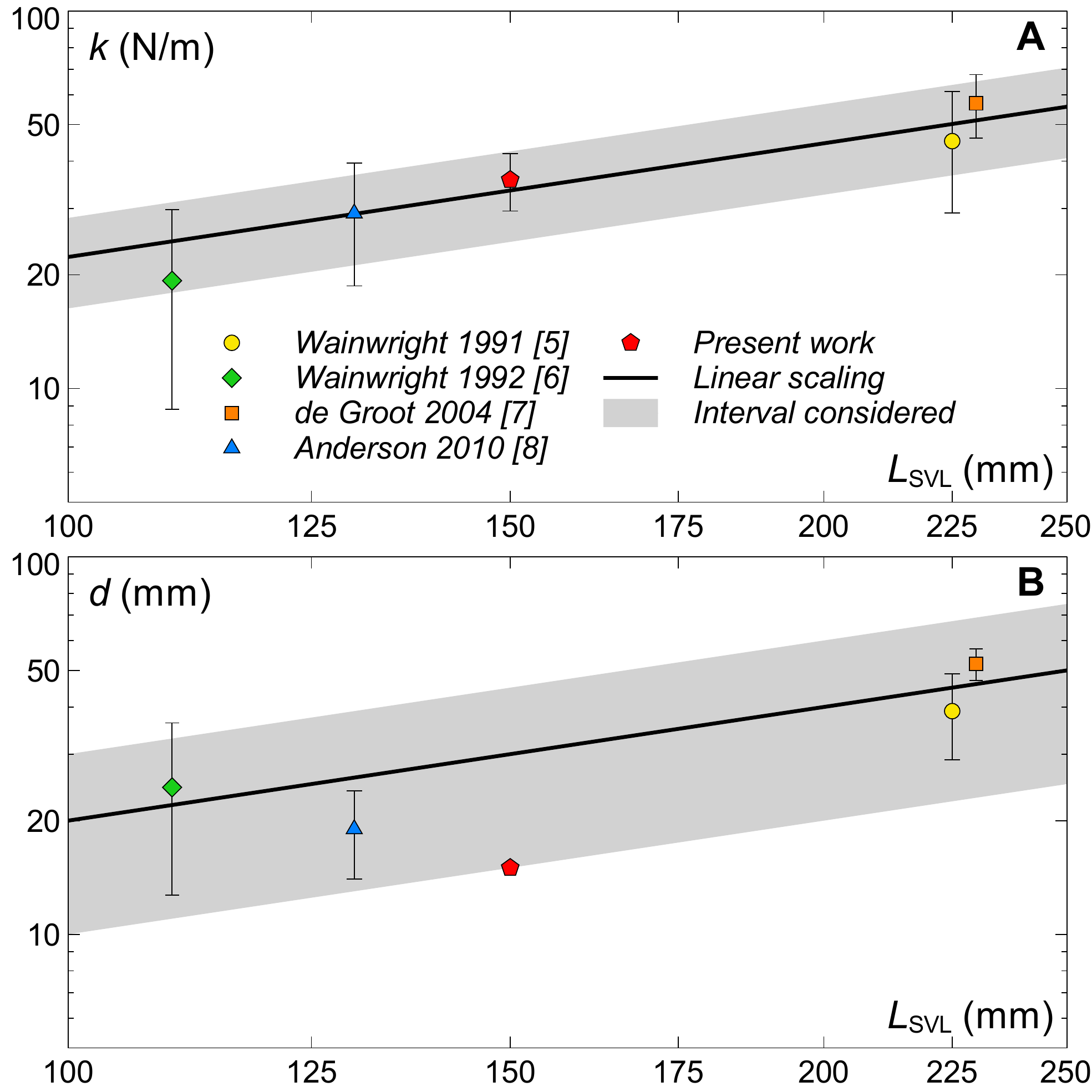}}
\caption{Evolution of the tongue stiffness ({\bf A}) and the parameter $d$ ({\bf B}) as a function of the SVL deduced from kinematics data of Refs.~[5-8].}
\label{fig-k}
\end{figure}

\noindent\textbf{\textit{Stiffness of the tongue pad}} -- The last morphological parameter needed in the main text is the stiffness of the tongue, $k$. According to the model developed in the main text, the maximum acceleration during the retraction phase is given by $a_{max}=k d/(m_t +m_p)$. Using $a_{max}$ and $d$ extracted from the kinematics data found in Refs.~[5-8], we calculated $k$ as shown in Fig.~\ref{fig-k}A. The corresponding values of $d$ are given in Fig.~\ref{fig-k}B. In good accuracy, $k$ and $d$ behaves linearly with $L_{\text{SVL}}$ as follows
\begin{align}
\label{k-tongue}
k &= (223 \pm 60)\, L_{\text{SVL}}, \\
d &= (0.2 \pm 0.1) \, L_{\text{SVL}},
\end{align}
in MKS units.

\begin{figure}[t]
\centerline{\includegraphics[width=\columnwidth]{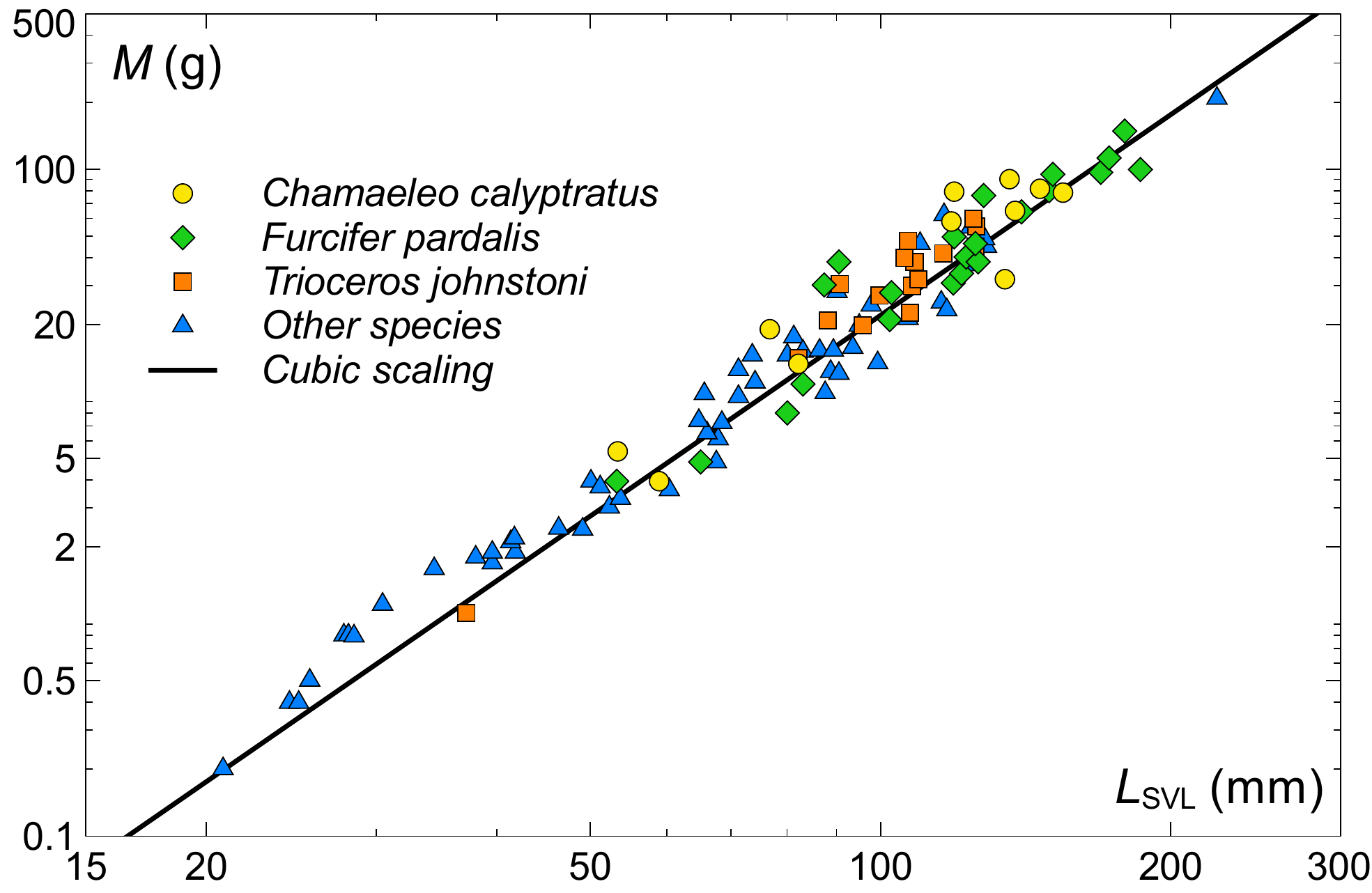}}
\caption{Evolution of the chameleon body mass as a function of the SVL for 105 chameleons among 23 species from Ref.~[2].}
\label{fig-mass}
\end{figure}

\noindent\textbf{\textit{Total body mass for chameleons}} -- In the main text, an estimation of the maximum prey size captured by viscous adhesion, $(V^*)^{1/3}$, has been obtained as a function of the chameleon sizes. The corresponding prey mass, $m^* = \rho V^*$, is then compared to the total mass of the chameleon, $M$. For this purpose, the evolution of $M$ with respect to $L_{\text{SVL}}$ should be known. This evolution is reported in Fig.~\ref{fig-mass} for a large number of specimens among various species~[2]. As expected, $M$ behaves as the third power of $L_{\text{SVL}}$ as follows
\begin{equation}
\label{mass-chameleon}
M = (21.9 \pm 4.0) \, L_{\text{SVL}}^3,
\end{equation}
in MKS units.

\begin{figure}[b]
\centerline{\includegraphics[width=\columnwidth]{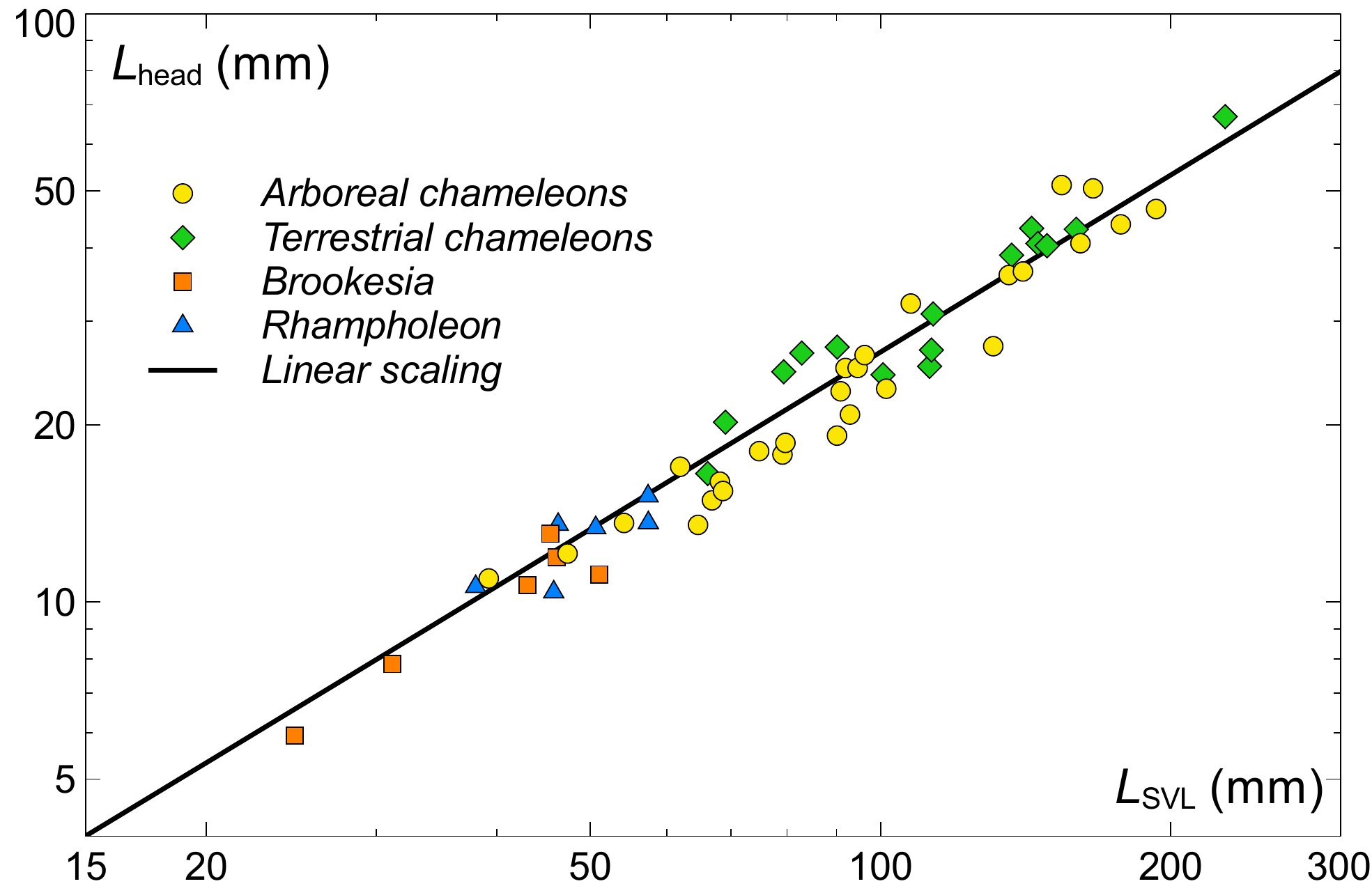}}
\caption{Evolution of the head length as a function of the SVL for 54 chameleons among several species~[10].}
\label{fighead}
\end{figure}

\noindent\textbf{\textit{Head length}} -- In order to compare the predictions of the model with available data, the measures performed in Ref.~[9] needed to be manipulated. In particular, the head length, $L_{\text{head}}$, must be related to the SVL. This evolution is shown in Fig.~\ref{fighead} 
for a significant number of specimens among several species~[10]. In good approximation, $L_{\text{head}}$ behaves linearly with $L_{\text{SVL}}$ as follows
\begin{equation}
\label{head-svl}
L_{\text{head}} = (0.27 \pm 0.05) \, L_{\text{SVL}}. 
\end{equation}

\begin{figure}[t]
\centerline{\includegraphics[width=\columnwidth]{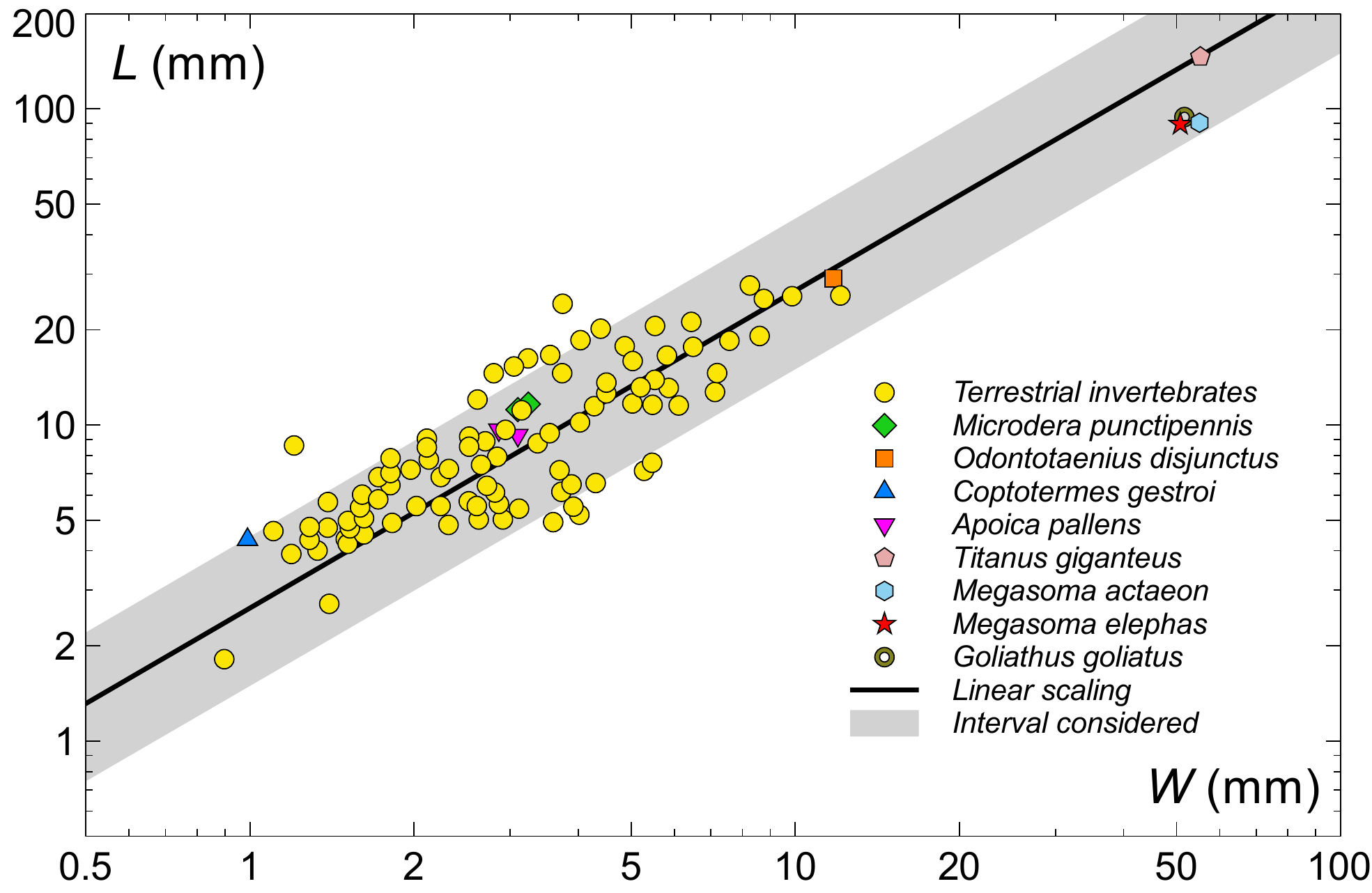}}
\caption{Evolution of the length with respect to the width for various invertebrates and insects as indicated (potential preys for chameleons). These data come from Refs.~[12-17]. Terrestial invertabrates include specimens from the following orders: Coleoptera, Homoptera, Hymenoptera, Orthoptera, Araneae and Microcoryphia. The length and width of \textit{Apoica pallens} are obtained from an average over 50 specimens (queens and workers)~[16]. Data for the largest Coleoptera are obtained from measurements on a photography found in Ref.~[17].}
\label{fig-inv}
\end{figure}

\section{Maximum prey size vs SVL - data manipulations}

In order to test the model presented in the main text, we compare its predictions with reported {\it in vivo} data giving the maximum prey size as a function of the chameleon size~[9,11]. The authors use the analysis of the stomachal contents for a large numbers of chameleons to determine various parameters including the mean maximum prey size. In Ref.~[11], the authors measured the maximum prey volume as a function of the SVL. These data can therefore be used directly in Fig.~4 of the main text. However, in Ref.~[9], the authors measured the maximum prey width as a function of the chameleon head length. The relation between the head length and the SVL has been obtained in Eq.~(\ref{head-svl}). Therefore, the complete conversion of the data requires the estimation of the prey volume from the prey width in order to obtain the maximum prey size as a function of SVL for both studies. 

The prey volume is approximated by the one of a cylinder: $V_{\text{prey}} = (\pi/4) W^2 L$, where $W$ is the width and $L$ the length of the prey. The evolution of $L$ with respect to $W$ is reported in Fig.~\ref{fig-inv} for a large set of invertebrates and insects~[12-17]. By using the following relation,
\begin{equation}
L = (2.8 \pm 1.5)\, W,
\end{equation}
90\% of the data are included (see grey area in Fig.~\ref{fig-inv}). This last relation allows to express the volume as a function of the width only. Therefore, the conversion for the prey size is made with the relation 
\begin{equation}
V_{\text{prey}}^{1/3} = (1.3 \pm 0.2) \, W.
\end{equation}

\section{Cavitation}
\label{sec:cavitation}

It has been previously shown that cavitation could affect the adhesion strength during the retraction of a probe connected to a fixed substrate through a viscous layer~[18]. 
Cavitation occurs if the depression in the fluid induced by the flow, $\Delta p=(3/\pi) {\eta \Omega \dot{h}}h^{-4}$, is larger than the atmospheric pressure: $\Delta p \simeq 2 f_a/\Sigma \gtrsim p_{\text{atm}}$, where $\Sigma=\Omega/h_0$ is the area of contact between the tongue and the prey~[18]. However, the viscous force is never larger than the retraction force, $f_a \le f_t$. The maximum retraction force is given by: $f_{max} = k d = (45\pm 25)\, L_{\text{SVL}}^2$ in good agreement with direct force measurements~[19]. Therefore, using Eq.~(\ref{sigma}), the depression in the fluid can never exceed $\Delta p \simeq 2 f_t/\Sigma \simeq (1.86 \pm 1.15)\, 10^{4}$ which is at least 3 times smaller than the atmospheric pressure. The retraction force is thus too small to allow the formation of cavitation.

\section{General equations of motion}

\begin{figure}[t]
\centerline{\includegraphics[width=0.8\columnwidth]{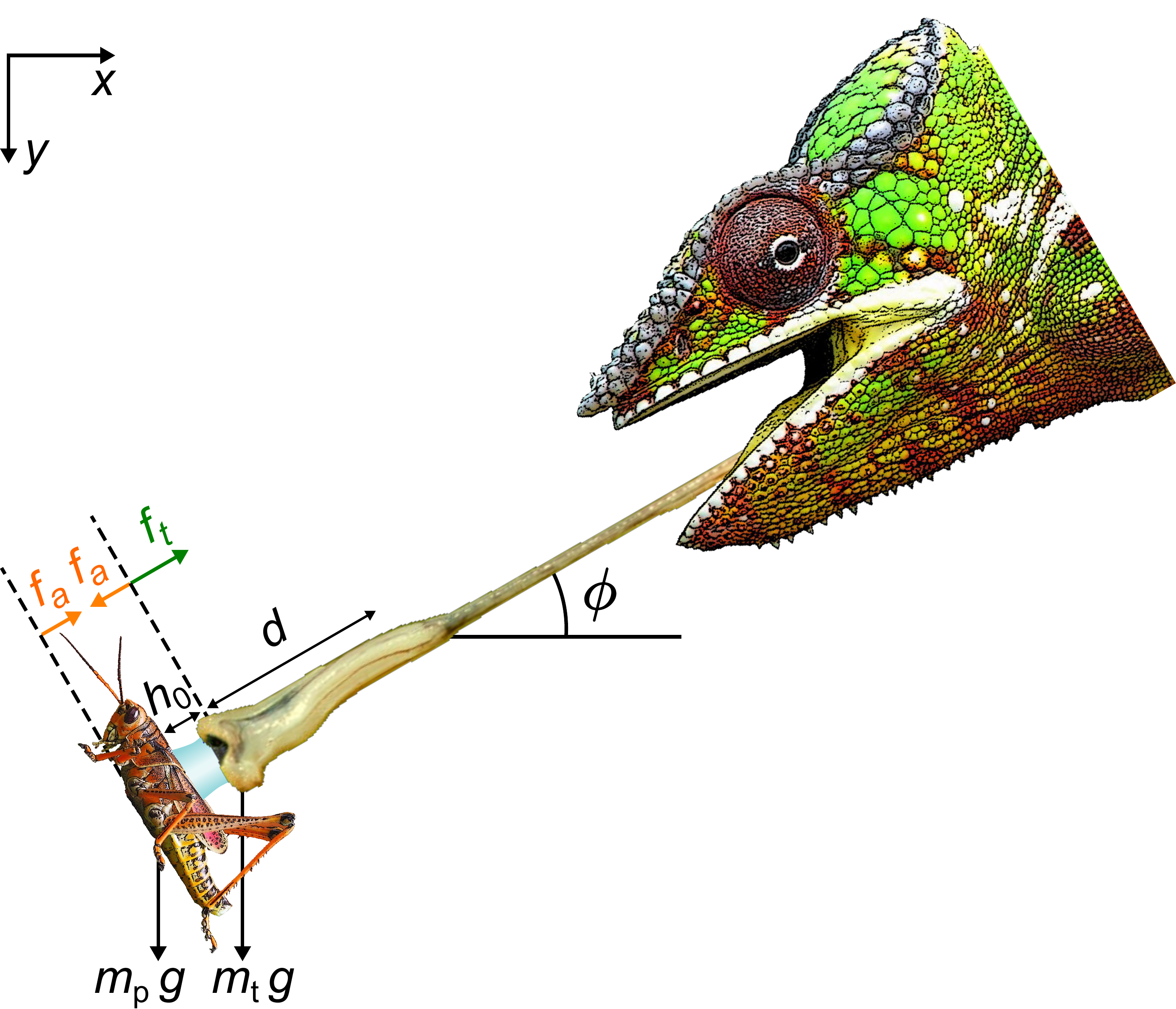}}
\caption{Schematic of the initial state of the retraction phase with the adhesion force $f_a$, the applied force $f_t$, the initial thickness of the mucus layer $h_0$ and the distance $d$ over which the retraction force applies along the direction $\phi$.}
\label{fig-schema}
\end{figure}

In this section, we derive the equations of motion taking into account gravity and the orientation $\phi$ of the retraction force. The retraction phase is described by considering a prey of mass $m_p$ at a position $(x_p, y_p)$ attached by a viscous fluid of thickness $h$ to a tongue of mass $m_t$ and position $(x_t,y_t)$. A force $f_t$ retracts the tongue with an angle $\phi$ with respect to the horizontal $x$-axis and produces an increase of the mucus thickness inducing a viscous force $f_a$ acting on both the tongue and the prey, see Fig.~\ref{fig-schema}. The equations of motion obtained by projecting the forces along $x$ and $y$-axis are
\begin{subequations}
\begin{align}
\label{xt-sup}
m_t \ddot{x}_t &= (f_t-f_a) \cos \phi, \\
\label{yt-sup}
m_t \ddot{y}_t &= (f_a-f_t)\sin \phi + m_t g, \\
\label{xp-sup}
m_p \ddot{x}_p &= f_a \cos \phi, \\ 
\label{yp-sup}
m_p \ddot{y}_p &= -f_a \sin \phi + m_p g.
\end{align}
\end{subequations}
Subtracting Eqs.~(\ref{xt-sup}) and (\ref{xp-sup}), we obtain
\begin{equation}
\ddot{h} = \frac{f_t}{m_t} - \left(\frac{1}{m_t}+\frac{1}{m_p}\right) f_a,
\end{equation}
where $h=(x_t-x_p)/\cos \phi = (y_p-y_t)/\sin \phi$. Subtracting Eqs.~(\ref{yt-sup}) and (\ref{yp-sup}) yields the same result. Therefore, the evolution of the mucus thickness $h$ depends neither on $g$ nor on $\phi$. We can thus consider $g=\phi=0$, as in the main text, to obtain the condition under which a capture is successful.

\section{Additional information about the model}

\subsection{Short time evolution}

Figures~3(c)-(d) of the main text show the evolutions of $H$ and $F_a$ as a function of $T$. At the onset of retraction (phase I), when $T\ll 1$, the tongue position is still close to its initial value, $X_t\simeq H_0$, as well as the variation of thickness, $\dot{H}\ll 1$. Therefore, Eq.~(3c) of the main text simplifies as $\ddot{H}\simeq D$ leading to $H\simeq H_0+D T^2/2$ and $F_a \simeq (D/H_0^5)T$. Since $D/H_0^5 \gg 1$, $F_a$ increases rapidly while $X_t\ll D$ leading to a decrease of $\ddot{H}$ given by $\ddot{H}\simeq D-[(1+m)/m]\dot{H}/H^{5}$. When $F_a$ reaches a value close to $mD/(1+m)$, $\ddot{H}\ll 1$ and stays small until $H$ diverges.
When $X_t$ is no longer negligible compared to $D$ while being smaller than $D$ and $\ddot{H}\ll 1$ (phase II), the adhesive force satisfies the equation
\begin{equation}
F_a \simeq F_a^{\text{app}} = m(D+H_0-X_t)/(1+m).
\end{equation}
This linear relationship between $F_a$ and $X_t$ compares well with the numerical solution as shown in Fig.~3(b) of the main text.

\subsection{Maximum prey mass}

Equation~(9) of the main text can also be obtained without solving the equations of motion. The work performed by the retraction force is given by $W= \int_0^{d} f_t dx = kd^{2}/2$. The maximum kinetic energy of the prey is obtained by integrating once Eq.~(2b) of the main text: $U^{(p)}_{max} = (\int_0^{t_d} f_a dt)^2/(2m_p)=\alpha^2/(32 m_p h_0^8 \Lambda^2)$, where we have used the explicit expression of the adhesive force and the saturation value of $h$. Equating these two last expressions with $\Lambda=1$ yields Eq.~(9) of the main text. Therefore the maximum prey mass can be estimated from the following relation
\begin{equation}
\label{general-mass}
m_p^{\star}= \frac{9}{128\pi^2} \frac{\eta^2 \Sigma^4}{h_0^4 W}, \quad W=\int_0^d f_t\, dx
\end{equation}
where $d$ is the distance over which the retraction force is significant. This relation may be useful if one considers a more sophisticated expression for the retraction force.

\section{Constant retraction force}

In this section, we consider the case of a constant retraction force $f_t$ applying on the tongue over a distance $d$ in order to show that the maximum prey size obtained in the main text does not depend significantly on the spatial variation of the force. Only the typical magnitude of the retraction force and the distance $d$ are important. The equations of motion are
\begin{subequations}
\label{model-sup}
\begin{align}
m_t \ddot{x}_t &= f_t\, {\cal H}(d+h_0-x_t)- \alpha \dot{h}h^{-5}, \\
\label{prey-eq-sup}
m_p \ddot{x}_p & = f_a =\alpha \dot{h}h^{-5}, \quad \alpha =3\eta \Sigma^2 h_0^2/2\pi, \\
h &= x_t-x_p,
\end{align}
\end{subequations}
with the initial conditions $x_p(0)= \dot{x}_p(0)= \dot{x}_t(0)=0$ and $x_t(0)=h_0$. This system is characterized by the length scale $\ell=(\alpha^2/m_t f_t)^{1/9}$ and the time scale $\tau=(\alpha\, m_t^4/f_t^5)^{1/9}$ which are used to adimensionalize the equations of motion as follows
\begin{subequations}
\label{eq-motion-sup}
\begin{align}
\label{tongue-sup}
\ddot{X}_t &= {\cal H}(D+H_0-X_t)- \dot{H} H^{-5}, \\
\label{prey-sup}
\ddot{X}_p &= m^{-1} \, \dot{H} H^{-5}, \quad m = m_p/m_t, \\
\label{H-sup}
H &= X_t-X_p.
\end{align}
\end{subequations}
The critical value $m^{\star}$ above which the prey detaches from the tongue is obtained when $H$ diverges precisely when $X_t = D+H_0$ (for $m>m^{\star}$ the detachment occurs for $X_t < D+H_0$). Therefore, to derive $m^{\star}$, it is enough to consider the equation in the interval $H_0 \le X_t\le D+H_0$ where the equation for the thickness evolution and the tongue position read
\begin{subequations}
\label{eq-motion-sup2}
\begin{align}
\label{tongue-sup2}
\ddot{X}_t &= 1- \dot{H} H^{-5}, \\
\label{thickness}
\ddot{H} &= 1-\frac{1+m}{m} \frac{\dot{H}}{H^{5}},
\end{align}
\end{subequations}
with $X_t(0)=H_0$, $\dot{X}_t(0)=0$, $H(0)=H_0$ and $\dot{H}(0)=0$. 

\begin{figure}[t]
\centerline{\includegraphics[width=\columnwidth]{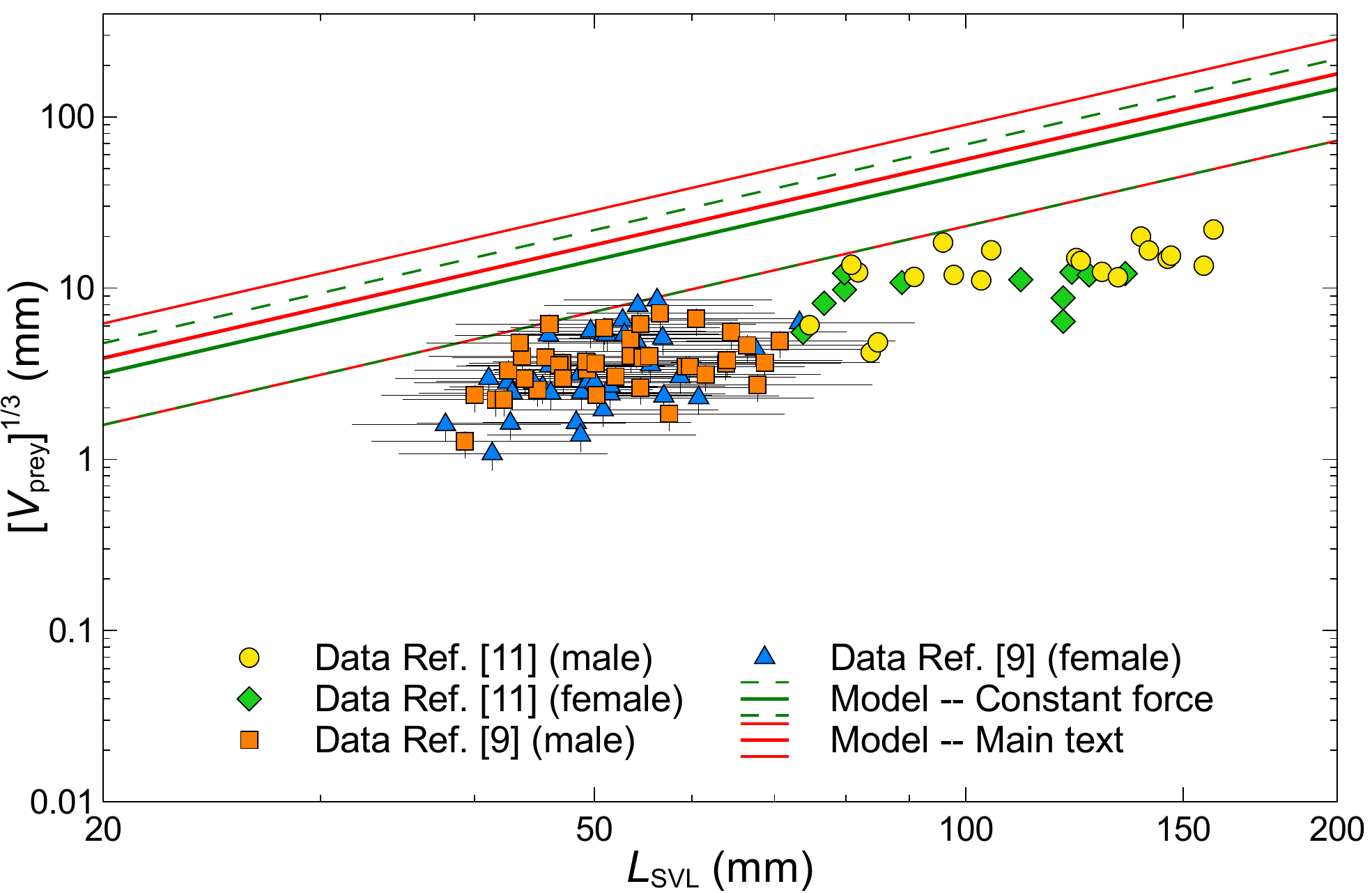}}
\caption{Measured maximum prey size as a function of $L_{\text{SVL}}$ obtained for a large sampling of chameleons~[9,11] plotted together with the predictions of a pure viscous adhesion model described either by Eq.~(10) of the main text or by Eq.~(\ref{prey-size}) (constant force).}
\label{figmass-sup}
\end{figure}

At the onset of retraction, when $T\ll 1$, the thickness is still close to its initial value, $\dot{H}\ll 1$. Therefore, Eq.~(\ref{thickness}) simplifies as $\ddot{H}\simeq 1$ leading to $H\simeq H_0+T^2/2$ and $F_a \simeq T/H_0^5$. Since $H_0^{-5} \gg 1$, $F_a$ increases rapidly leading to a decrease of $\ddot{H}$. When $F_a$ reaches a value close to $m/(1+m)$, $\ddot{H}\ll 1$ and stays small until $H$ diverges. In good approximation, the adhesive force thus satisfies the equation
\begin{equation}
\label{fa}
F_a \simeq F_a^{\text{app}}=\dot{H} H^{-5} = m/(1+m).
\end{equation}
An integration of Eqs.~(\ref{tongue-sup2}) and (\ref{fa}) leads to the evolution of the tongue position and of the mucus thickness as a function of time
\begin{subequations}
\label{solution}
\begin{align}
\label{Xt-app}
X_t &\simeq H_0 + \frac{T^2}{2(1+m)}, \\
\label{H-app-sup}
H &\simeq H^{\text{app}} = H_0\left[\frac{\Lambda}{\Lambda-T}\right]^{1/4}, \quad \Lambda = \frac{1+m}{4 m H_0^4}.
\end{align}
\end{subequations}
The thickness $H$ is thus seen to diverge at time $T^{\star}=\Lambda$. The critical mass $m^{\star}$ is obtained from the condition  $X_t(T^{\star})=D + H_0$. From Eq.~(\ref{Xt-app}), the definition of $\Lambda$ and assuming $m^{\star} \gg 1$, we obtain
\begin{equation}
m^{\star}=\frac{1}{32 D H_0^8}.
\end{equation}
Returning to the physical variables, the maximum prey mass is therefore given by
\begin{equation}
\label{masse-sup}
m_p^{\star} = \frac{9}{128\pi^2} \frac{\eta^2 \Sigma^4}{f_t\, d\, h_0^4}.
\end{equation}
Notice that this expression coincide precisely with Eq.~(\ref{general-mass}) when $f_t$ is constant. Using the parameter values reported in the main text together with $f_t = f_{max}$ obtained in Section~\ref{sec:cavitation}, Eq.~(\ref{masse-sup}) can be written as a function of the chameleon body size as
\begin{equation}
\label{prey-size-sup}
{V^{\star}}^{1/3} = (2.2 \pm 1.1)\, L_{\text{SVL}}^{5/3},
\end{equation}
in MKS units. Figure~\ref{figmass-sup} shows a comparison between data from Refs.~[9,11], Eq.~(10) of the main text and Eq.~(\ref{prey-size-sup}). We thus see that, provided $f_t$ has the correct order of magnitude, both models yield similar results.

\begin{center}
\line(1,0){250}
\end{center}

\begin{enumerate}[label={[\arabic*]},nolistsep]
\item J. Bico, J. Ashmore-Chakrabarty, G. H. McKinley and H. A. Stone, \textit{Phys. Fluids} {\bf 21}, 082103 (2009).
\item C. V. Anderson, T. Sheridan and S. M. Deban, \textit{J. Morphol.} {\bf 273}, 1214 (2012). See also their supplementary document containing all the data.
\item J. Mendez and A. Keys, \textit{Metabolism} {\bf 9}, 184 (1960).
\item A. Herrel, J. J. Meyers, P. Aerts and K. C. Nishikawa, \textit{J. Exp. Biol.} {\bf 203}, 3255 (2000).
\item P. C. Wainwright, D. M. Kraklau, and A. F. Bennett, \textit{J. Exp. Biol.} {\bf 159}, 109 (1991).
\item P. C. Wainwright and A. F. Bennett, \textit{J. Exp. Biol.} {\bf 168}, 1 (1992).
\item J. H. de Groot and J. L. van Leeuwen, \textit{Proc. R. Soc. Lond. B} {\bf 271}, 761 (2004).
\item C. V. Anderson and S. M. Deban, \textit{Proc. Natl. Acad. Sci.} {\bf 107}, 5495 (2010).
\item G. J. Measey, A. D. Rebelo, A. Herrel, B. Vanhooydonck and K. A. Tolley, \textit{J. Zool.} {\bf 285}, 247 (2011).
\item R. Bickel and J. B. Losos, \textit{Biol. J. Linnean Soc.} {\bf 76}, 91 (2002).
\item F. Kraus, A. Medeiros, D. Preston, C. E. Jarnevich and G. H. Rodda, \textit{Biol. Invasions} {\bf 14}, 579 (2012).
\item J. L. Sabo, J. L. Bastow and M. E. Power, \textit{J. N. Am. Benthol. Soc.} {\bf 21}, 336 (2002).
\item Y. Wang, X. Liu, J. Zhao, K. Rexili and J. Ma, \textit{J. Insect Sci.} {\bf 11}, article 39 (2011).
\item A. K. Davis, B. Attarha and T. J. Piefke, \textit{J. Insect Sci.} {\bf 13}, article 107 (2013).
\item A. M. Costa-Leonardo, A. Arab and F. E. Casarin, \textit{J. Insect Sci.} {\bf 4}, article 10 (2004).
\item R. L. Jeanne, C. A. Graf and B. S. Yandell, \textit{Naturwissenschaften} {\bf 82}, 296 (1995).
\item D. M. Williams, Book of insect records, University of Florida, chap. 30\\ (http://entnemdept.ufl.edu/walker/ufbir/index.shtml)
\item S. Poivet, F. Nallet, C. Gay, J. Teisseire and P. Fabre, \textit{Eur. Phys. J. E} {\bf 15}, 97 (2004).
\item A. Herrel, J. J. Meyers, P. Aerts and K. C. Nishikawa, \textit{J. Exp. Biol.} {\bf 204}, 3621 (2001).
\end{enumerate}


\begin{thebibliography}{99}

\bibitem{schwenk} 
K. Schwenk, \textit{Feeding: Form, Function, and Evolution in Tetrapod Vertebrates} (Academic Press, San Diego, 2000).

\bibitem{wainwright91}
P. C. Wainwright, D. M. Kraklau and A. F. Bennett, \textit{J. Exp. Biol.} {\bf 159}, 109 (1991).

\bibitem{wain92a}
P. C. Wainwright and A. F. Bennett, \textit{J. Exp. Biol.} {\bf 168}, 1 (1992).

\bibitem{wain92b}
P. C. Wainwright and A. F. Bennett, \textit{J. Exp. Biol.} {\bf 168}, 23 (1992).
 
\bibitem{degroot04}
J. H. de Groot and J. L. van Leeuwen, \textit{Proc. R. Soc. Lond. B} {\bf 271}, 761 (2004).

\bibitem{herr09}
A. Herrel, S. M. Deban, V. Schaerlaeken, J.-P. Timmermans and D. Adriaens, \textit{Physiol. Biochem. Zool.} {\bf 82}, 29 (2009).

\bibitem{higham13}
T. E. Higham and C. V. Anderson, in \textit{The Biology of Chameleons}, edited by K. A. Tolley and A. Herrel (University of California Press, Oakland, 2013), p. 63.

\bibitem{herrel00} 
A. Herrel, J. J. Meyers, P. Aerts and K. C. Nishikawa, \textit{J. Exp. Biol.} {\bf 203}, 3255 (2000). 

\bibitem{stefan}
J. Stefan, \textit{Sitzungsber. Akad. Wiss. Wien: Math. Naturwiss. Kl.} {\bf 69}, 713 (1874).

\bibitem{bikerman}
J. J. Bikerman, \textit{J. Colloid Sci.} {\bf 2}, 163 (1947).




\bibitem{bico09} 
J. Bico, J. Ashmore-Chakrabarty, G. H. McKinley and H. A. Stone, \textit{Phys. Fluids} {\bf 21}, 082103 (2009).


\bibitem{rem-gamma}
We have used a conservative estimation for the surface tension very close to the one of water. Smaller values of $\gamma$ leads to even larger value of the viscosity reinforcing our conclusion. Notice however that small variations of $\gamma$ has little impact on the viscosity since $\eta\sim 1/\gamma^{0.35}$.

\bibitem{briedis}
D. Briedis, M. F. Moutrie and R. T. Balmer, \textit{Rheol. Acta} {\bf 19}, 365 (1980).

\bibitem{harkness} 
L. Harkness, \textit{Nature} {\bf 267}, 346 (1977).

\bibitem{mull04}
U. K. M\"uller and S. Kranenbarg, \textit{Science} {\bf 304}, 217 (2004)

\bibitem{anderson10}
C. V. Anderson and S. M. Deban, \textit{Proc. Natl. Acad. Sci.} {\bf 107}, 5495 (2010).

\bibitem{herrel01} 
A. Herrel, J. J. Meyers, P. Aerts and K. C. Nishikawa, \textit{J. Exp. Biol.} {\bf 204}, 3621 (2001).

\bibitem{sup_mat} See Supplemental Material.

\bibitem{rem1}
The amplitude of this force could be lowered if cavitation occurs. However, the retraction force obtained by direct measurements ($\lesssim 1$ N) is too small by at least a factor three to allow the formation of cavitation~\cite{sup_mat}.

\bibitem{rem2}
Notice that gravity or the direction of the applied force with respect to the horizontal have no influence on the evolution of $h(t)$ which determines the success of a capture~\cite{sup_mat}.

\bibitem{matt08}
P. G. D. Matthews and R. S. Seymour, \textit{J. Exp. Biol.} {\bf 211}, 3790 (2008).

\bibitem{rem3}
With these values of the parameters and considering an ordinary prey mass weighing 1\% of the chameleon mass and $L_{\text{SVL}}=0.15$ m, we find that the maximum retraction force and acceleration are given by $f_{max}= kd \simeq (0.95\pm 0.55)$ N and $a_{max} = kd/(m_t+m_p) \simeq (420 \pm 240)$ m/s$^2$. The time needed for the tongue to retract by a distance $d$ reads $t_{d} =\pi \tau/(2\omega) \simeq (13\pm 2)$ ms whereas the constant velocity of the tongue and the prey beyond that point is $v_{d}=d\omega/\tau \simeq (3.5\pm 1.8)$ m/s in good agreement with force measurements~\cite{herrel01} and typical kinematics data (see Fig.~\ref{figcine} and Refs.~\cite{wainwright91,wain92a,degroot04,anderson10}).


\bibitem{measey11}
G. J. Measey, A. D. Rebelo, A. Herrel, B. Vanhooydonck and K. A. Tolley, \textit{J. Zool.} {\bf 285}, 247 (2011).

\bibitem{kraus12}
F. Kraus, A. Medeiros, D. Preston, C. E. Jarnevich and G. H. Rodda, \textit{Biol. Invasions} {\bf 14}, 579 (2012).




\bibitem{deban97}
S. M. Deban, D. B. Wake and R. Roth, \textit{Nature} {\bf 389}, 27 (1997).

\bibitem{deban06}
S. M. Deban, J. C. O'Reilly, U. Dicke and J. L. van Leeuwen, \textit{J. Exp. Biol.} {\bf 210}, 655 (2006).



\end{thebibliography}
\end{document}